\documentclass{article}
\usepackage{spconf,amsmath,graphicx}
\usepackage{epsfig,amssymb,booktabs,multirow,enumitem,caption,setspace}

\title{SG-JND: Semantic-Guided Just Noticeable Distortion Predictor For Image Compression}

\name{
 \parbox{\linewidth}{\centering
  Linhan Cao$^{1,2}$, Wei Sun$^{1*}$\thanks{$^*$Corresponding authors}, Xiongkuo Min$^1$, Jun Jia$^1$, Zicheng Zhang$^1$, Zijian Chen$^1$,\\ Yucheng Zhu$^1$, Lizhou Liu$^2$, Qiubo Chen$^2$, Jing Chen$^2$, Guangtao Zhai$^{1*}$\thanks{This work was supported in part by the China Postdoctoral Science Foundation under Grants 2023TQ0212 and 2023M742298, the Postdoctoral Fellowship Program of CPSF under Grant GZC20231618, and the National Natural Science Foundation of China under Grants 62301316, 62225112, 62101325 and 62101326.}
}}

  \address{$^1$Institute of Image Communication and Network Engineering, Shanghai Jiao Tong University \\
 $^2$Audio and Video Lab,  Xiaohongshu}

\begin{document}
%
\maketitle

\begin{abstract}
Just noticeable distortion (JND), representing the threshold of distortion in an image that is minimally perceptible to the human visual system (HVS), is crucial for image compression algorithms to achieve a trade-off between transmission bit rate and image quality. However, traditional JND prediction methods only rely on pixel-level or sub-band level features, lacking the ability to capture the impact of image content on JND. To bridge this gap, we propose a \textit{\underline{S}emantic}-\textit{\underline{G}uided} JND (SG-JND) network to leverage semantic information for JND prediction. In particular, SG-JND consists of three essential modules: the image preprocessing module extracts semantic-level patches from images, the feature extraction module extracts multi-layer features by utilizing the cross-scale attention layers, and the JND prediction module regresses the extracted features into the final JND value. Experimental results show that SG-JND achieves the state-of-the-art performance on two publicly available JND datasets, which demonstrates the effectiveness of SG-JND and highlight the significance of incorporating semantic information in JND assessment.
\end{abstract}
\begin{keywords}
Just noticeable distortion, deep learning, semantic analysis, human visual system
\end{keywords}

\begin{figure*}[t]
\centering
\centerline{\epsfig{figure=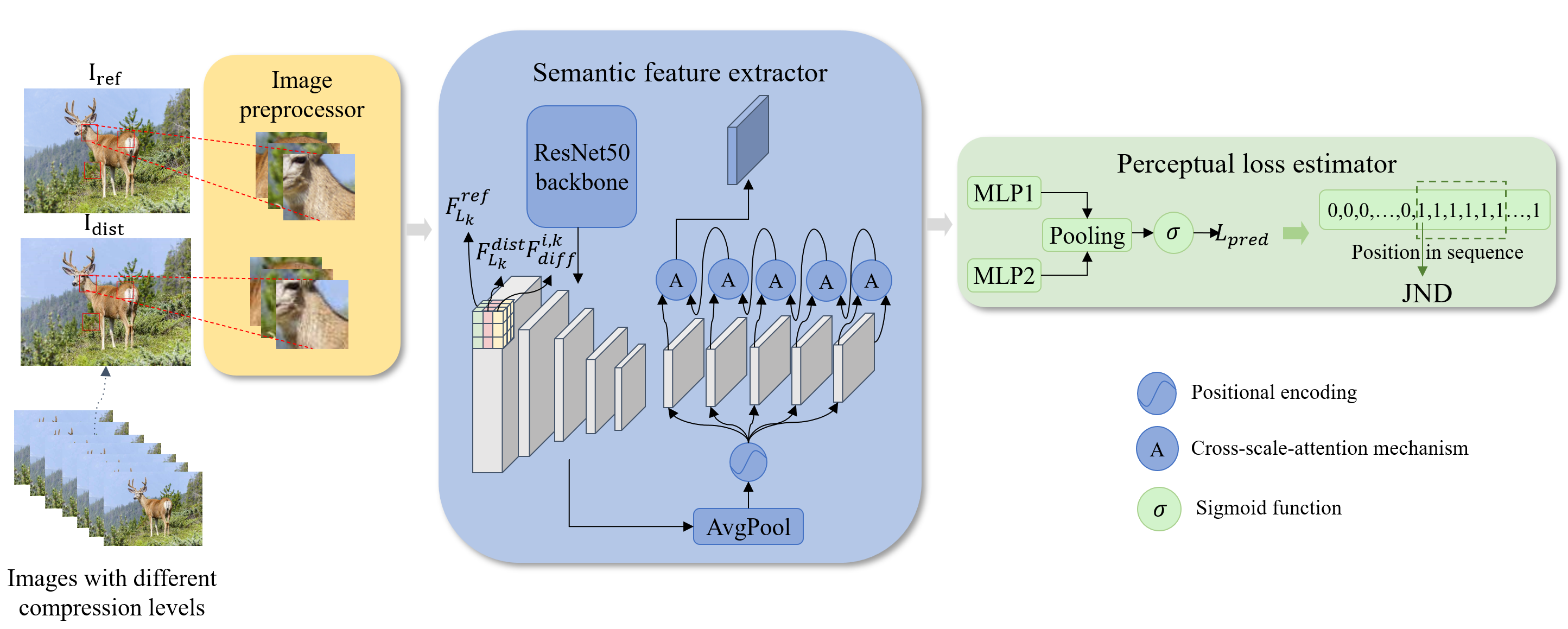,width=18cm}}
\caption{The overall framework of the proposed SG-JND model.}
\label{fig:image1}
\end{figure*}

\section{Introduction}
\label{sec:intro}

Just noticeable distortion (JND) is a concept in image processing \cite{sun2022deep, sun2024enhancing, sun2024analysis} that identifies the smallest change in an image that the human visual system (HVS) can perceive. This threshold is vital for optimizing image compression, as it presents an opportunity to reduce bandwidth and storage requirements without compromising perceived quality. Moreover, JND can be applied in discreet information embedding like watermarking, etc. 

For image compression, recent studies \cite{hu2016gmm,jin2016statistical,fan2019picture,zhang2021deep} have shown that within a certain range of bit rates, humans can discern only a limited number of distinct quality levels. Therefore, JND corresponds to the compressed quality level at which the human eye can just perceive the change in quality. This insight has led to the creation of JND-based image and video datasets, such as MCL-JCI \cite{jin2016statistical}, JND-Pano \cite{liu2018jnd}, KonJND-1k \cite{lin2022large}, and VideoSet \cite{wang2017videoset}, facilitating subjective visual quality studies. While these subjective assessments are reliable, they are time-consuming and labor-intensive. Consequently, a group of objective JND prediction methods have emerged to automatically predict the JND values without the involvement of human subjects.

Existing JND models fall into two primary categories: pixel-domain models \cite{liu2010,wu2013,wu2017} and sub-band domain models \cite{wei2009,bae2013,bae2016dct}. Pixel-domain models calculate JND for each pixel, focusing on background luminance adaptation and spatial contrast masking. Liu \textit{et al.} \cite{liu2010} employed a total-variation-based image decomposition algorithm to separate an image into structural and textural components for better contrast masking estimation. Wu \textit{et al.} \cite{wu2013} used an autoregressive model for orderly content prediction based on the free-energy principle, enhancing JND estimation by distinguishing between orderly and disorderly content. By incorporating pattern complexity and orientation selectivity, Wu \textit{et al.} \cite{wu2017} developed an improved JND model for images, significantly improving spatial masking effect estimation.

Sub-band domain models convert images into sub-bands for JND computation, emphasizing the Contrast Sensitive Function (CSF), luminance adaptation, contrast, etc. We \textit{et al.} \cite{wei2009} presented a spatio-temporal JND profile for grayscale DCT domain images and videos, improving estimation by integrating spatial and temporal CSF, luminance adaptation, and contrast masking, with considerations for retinal movement and motion directionality. Bae \textit{et al.} \cite{bae2013} introduced a DCT-based JND model, focusing on luminance adaptation in the DCT frequency domain, validated through psychophysical experiments. It highlighted how DCT frequency and background luminance crucially influence JND thresholds, which display quasi-parabolic patterns, mirroring the human visual system's response to luminance changes. Bae \textit{et al.} \cite{bae2016dct} developed a JND model that considers joint effects between temporal masking and foveated masking to elaborately estimate JND thresholds for various image regions.

However, both pixel and sub-band domain JND models assess each pixel or sub-band individually, potentially limiting their effectiveness in reflecting the JND threshold for the whole image. In recent years, deep learning-based methods have been applied to JND prediction. Tian \textit{et al.} \cite{tian2020just} proposed a CNN model to extract image features for JND level prediction and a support vector regression model to predict the JND level numbers. Liu \textit{et al.} \cite{liu2019PW-JND} treated JND prediction as a multi-classification problem, using a deep neural network that contained a perceptual lossy/non-lossy predictor and employed a sliding-window search strategy to accurately locate the value of the JND. Fan \textit{et al.} \cite{fan2019net} proposed a method for predicting the Satisfied User Ratio (SUR) curve, treating JND as a discrete variable and the SUR function as its complementary cumulative distribution function, utilizing a siamese network for image pair analysis. Building upon Fan \textit{et al.}'s model, Lin \textit{et al.} \cite{lin2020featnet} optimized SUR distribution parameters through maximum likelihood estimation and the Anderson-Darling test, and incorporated transfer and deep feature learning in a two-stage model to enhance JND prediction accuracy.

In this work, we introduce \textit{\underline{S}emantic}-\textit{\underline{G}uided} JND (SG-JND), a new JND prediction approach that fuses semantic guidance with multi-layer features to determine the JND values of images. Firstly, we divide boh the reference and distorted images into non-overlapping patches. Secondly, we extract the multi-layer features from a backbone network (i.e. ResNet-50 \cite{he2016deep}) and further employ a cross-scale attention (CSA) module to leverage high-level semantic information to guide multi-layer feature fusion, achieving the semantic-aware JND feature extraction. Thirdly, the JND prediction module evaluates and assigns weights to each patch to obtain a perceptual label for each distorted image, and then combines the labels of each distorted image to determine the final JND value, thus providing a comprehensive picture of the degradation of perceptual image quality. SG-JND demonstrates state-of-the-art performance in JND prediction on two public JND benchmarks, MCL-JCI and KonJND-1k.

\section{PROPOSED METHOD}
Our proposed SG-JND is structured into three parts: 1) image preprocessing module to randomly extract non-overlapping patches from images; 2) feature extraction module to extract JND features guided by high-level semantics; 3) JND prediction module to evaluate and weigh patch quality to accurately predict the JND value. The overall framework of the proposed SG-JND model is illustrated in Fig.\ref{fig:image1}.

\subsection{Image Processing Module}
Our image preprocessing module, denoted as $\text{P} _{patch} $, systematically divides the image into patches, facilitating dataset expansion.

In the training dataset, the reference image is the original image without distortion, and the distorted image is the original image compressed by a certain quality factor (QF). Taking JPEG format compression as an example, there are 100 different distortion levels corresponding to \( QF \) ranging from 1 to 100. We denoted a reference image as $ I_{ref} $ and its corresponding distorted ones as $ I_{dist} $ with  \( QF \) ranging from 1 to 100. For each image pair \( (I_{ref}, I_{dist}^{qf}) \), where \( qf \in [1, 100] \) and \( I_{dist}^{qf} \) represents the distorted image at a specific compression level, \( P_{patch} \) randomly crops \( N \) patches of size \( s \times s \) from identical locations on both the \(I_{ref}\) and the \( I_{dist}^{qf} \). This process is mathematically formulated as:

\begin{eqnarray}
P_{ref}^{i}, P_{dist}^{i,qf} = P_{patch}(I_{ref}, I_{dist}^{qf}, N, s),i = 1,2,3, \dots , N,
\end{eqnarray}

\noindent
where \( P_{ref}^{i} \) and \( P_{dist}^{i,qf} \) are the \( i^{th} \) patches from the reference and distorted images, respectively. These patch pairs then serve as the input for the feature extraction module, allowing for a more nuanced and detailed analysis than the full image.

\begin{figure}[h]
\centering
\centerline{\epsfig{figure=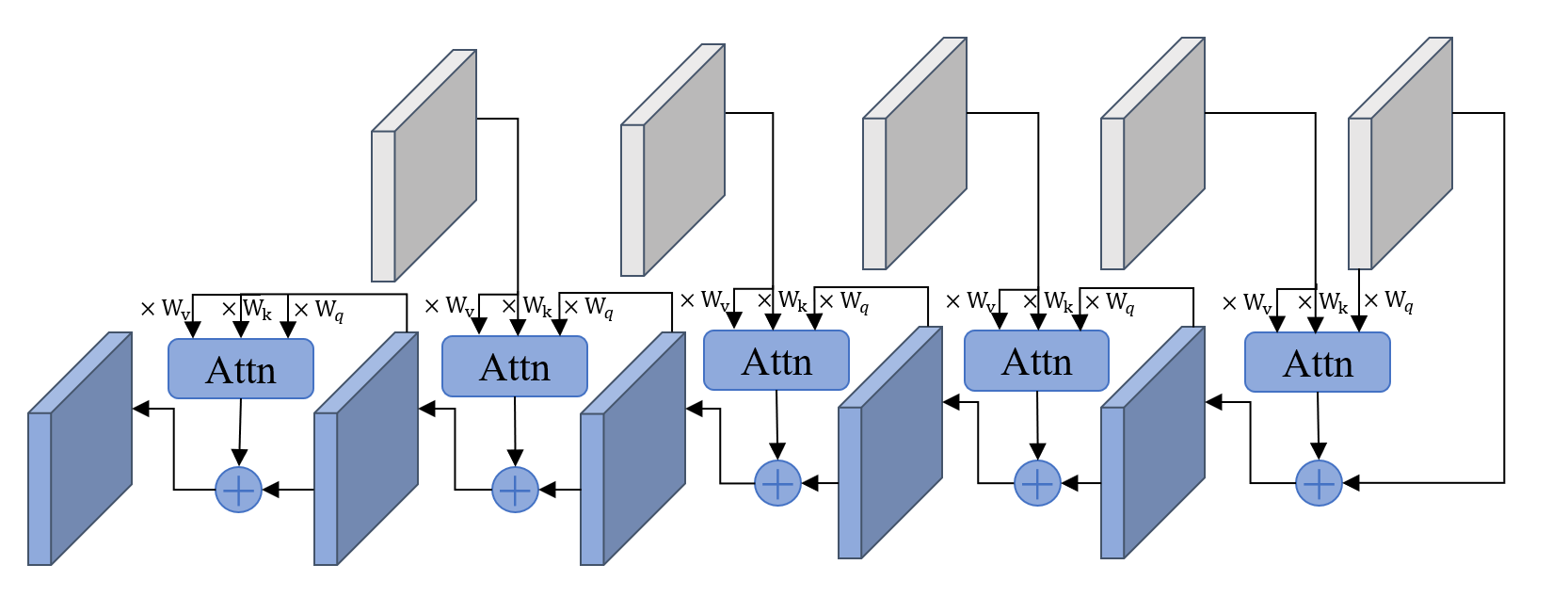,width=9cm}}
\caption{The structure of the cross-scale mechanism.}
\label{CSA}
\end{figure}

\subsection{Feature Extraction Module}

The feature extraction module, symbolized as \( F_{extract} \), is responsible for processing input patch pairs \( P_{ref}^{i} \) and \( P_{dist}^{i,qf} \). This module can be divided into several key steps:

\begin{enumerate} 
    \item Semantic feature extraction: Each patch pair is fed into a ResNet-50 model pre-trained on ImageNet. This step extracts multi-scale features, denoted as \( F_{L_k}^{ref}(P_{ref}^{i}) \) and \( F_{L_k}^{dist}(P_{dist}^{i}) \) for \( k = 1, 2, \dots, 5 \), representing the five stages from low to high. The deep features extracted from low to high layers capture a broad spectrum of semantic details, crucial for perceiving subtle image distortions.
    \item Semantic feature concatenation: We then compute the difference \( F_{diff}^{i, k} \) between reference and distorted patch features to highlight the semantic features at different levels of distortion, enhancing the model's sensitivity to JND. Concatenating the reference, distortion, and differential features for each layer and resizing them through average pooling to a uniform dimension aligns with the highest-level features to integrate semantic information of varying granularities and diminishes the computational complexity:

\begin{eqnarray}
F_{resized}^{i, k} = \text{AvgPool}(\text{cat}(F_{L_k}^{ref}, F_{L_k}^{dist}, F_{diff}^{i, k})),
\end{eqnarray}

    \item Cross-Scale Feature Processing: Inspired by the cross-scale attention mechanism in CFANet \cite{chen2024topiq} which can effectively propagate high-level semantic information from coarse to fine scales, our approach also use high-level features as queries to select semantic important low-level distortion features. The CSA mechanism is formulated as a query problem based on feature similarities, where high-level features serve as queries(Q), and low-level features form key(K) and value(V) pairs. We also add position encoding, denoted as PE, before CSA to provide location information for subsequent processing.
    
The detailed structure of CSA is illustrated in Fig. \ref{CSA}. CSA employs scaled dot-product attention \cite{vaswani2017attention}, defined by the following equation:

\begin{eqnarray}
\text{Attn}(Q, K, V) = softmax(\frac{QK^T}{\sqrt{d_k}})V,
\end{eqnarray}

where $d_k$ represents the feature dimension. The CSA mechanism can be expressed as follows:

\begin{equation}
	\begin{split}
F_{CSA}^{i, k} &= \text{CSA}((F_{PE}^{i, k}, F_{CSA}^{i+1, k}) \\
&= \text{Attn}(W_qF_{CSA}^{i+1, k}, W_kF_{PE}^{i, k}, W_vF_{PE}^{i, k}) + F_{CSA}^{i+1, k},
	\end{split}
\end{equation}


where $F_{PE}^{i, k} = F_{resized}^{i, k} + PE$ and $i= 1,2,3,4$. Finally, we obtain the finest-scale feature $F_{CSA}^{i, 1}$ that has been guided by high-level semantic features, which focus on the representation of semantic distortion features. The attention mechanism is applied once more to the obtained $F_{CSA}^{i, 1}$ itself which ensures that the distortion features at each position in the final feature vector are combined to more accurately reflect the degradation of the patch quality:

\begin{eqnarray}
F_{final}^{i} = \text{CSA} (F_{CSA}^{i, 1}, F_{CSA}^{i, 1}).
\end{eqnarray}

\end{enumerate}

Through integrating multi-layer semantic analysis with advanced attention mechanisms, SG-JND captures the nuances of human visual perception in images.

\begin{figure*}[t]
\hspace{2cm}
\begin{minipage}[b]{0.33\linewidth} 
  \centering
  \centerline{\includegraphics[width=6cm]{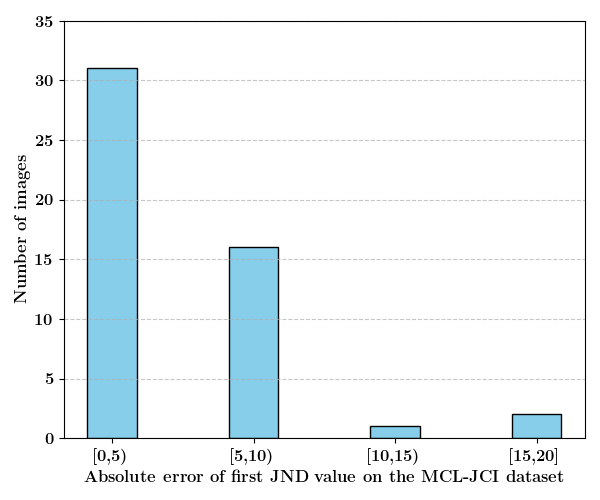}}
  \centerline{(a)}\medskip
\end{minipage}
\hspace{1cm} 
\begin{minipage}[b]{0.33\linewidth} 
  \centering
  \centerline{\includegraphics[width=6cm]{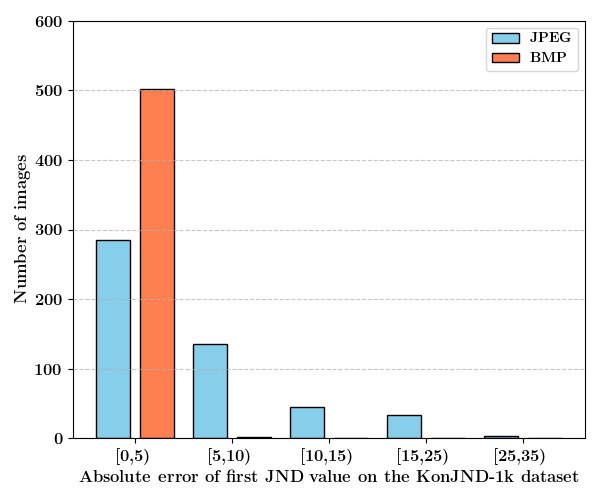}}
  \centerline{(b)}\medskip
\end{minipage}
\caption{Statistics of experimental results on the two datasets. (a)Histogram of the absolute error between the predicted JND and the ground truth JND on the MCL-JCI dataset. (b)Histogram of the absolute error between the predicted JND and the ground truth JND on the KonJND-1k dataset.}
\label{histogram}
\end{figure*}

\subsection{JND Prediction Module}
The JND prediction module, denoted as \( J_{pred} \), determines the perceptual threshold at which distortions within an image transition from imperceptible to perceptible.

\begin{figure*}[t]

\begin{minipage}[b]{0.33\linewidth}
  \centering
  \centerline{\includegraphics[width=6.0cm]{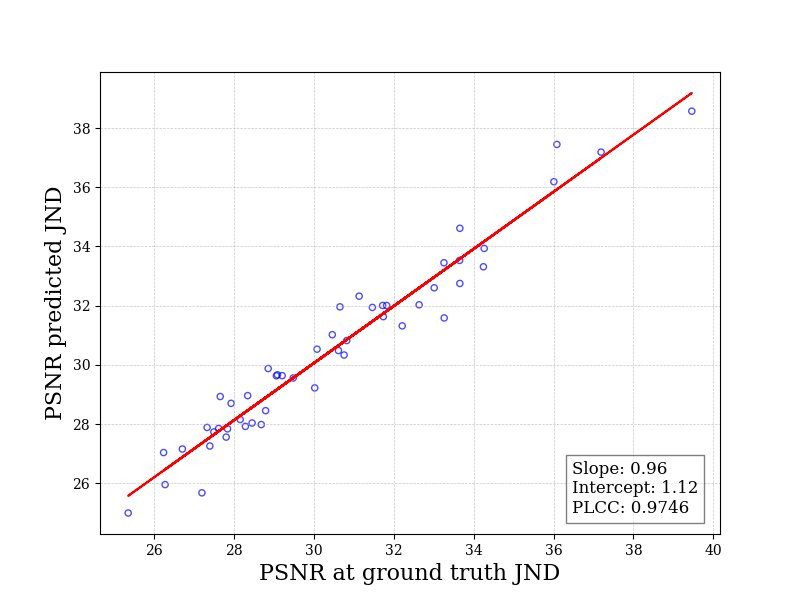}}
  \centerline{(a) MCL-JCI}\medskip
\end{minipage}
\begin{minipage}[b]{0.33\linewidth}
  \centering
  \centerline{\includegraphics[width=6.0cm]{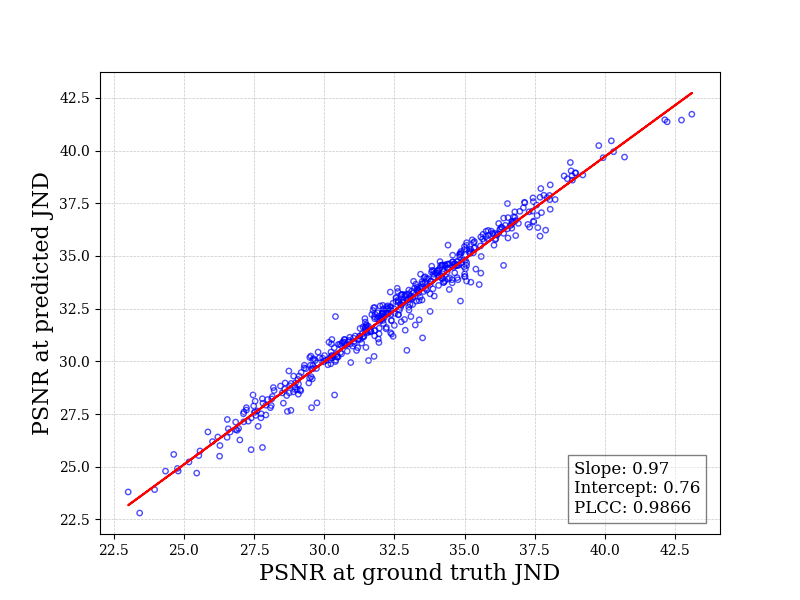}}
  \centerline{(b) konJND-1k(JPEG)}\medskip
\end{minipage}
\hfill
\begin{minipage}[b]{0.33\linewidth}
  \centering
  \centerline{\includegraphics[width=6.0cm]{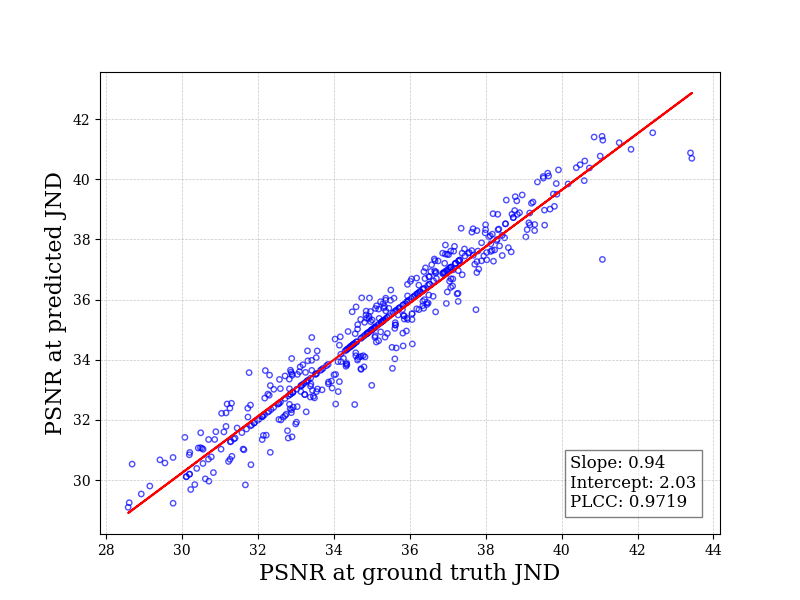}}
  \centerline{(c) konJND-1k(BMP)}\medskip
\end{minipage}
\caption{PSNR comparison between the ground truth JND images and predicted JND images on the MCL-JCI dataset (a), the JPEG compression format images of the KonJND-1k dataset (b), the BMP compression format images of the KonJND-1k dataset (c). The corresponding PLCCs are 0.9746, 0.9866, and 0.9719.}
\label{plcc}
\end{figure*}

\begin{table*}[t]
\begin{center}
\captionsetup{font={footnotesize}}
\fontsize{8pt}{10pt}\selectfont
\caption{PERFORMANCE RESULTS OF THE PROPOSED MODEL AND COMPARED METHODS ON TWO JND DATASETS.}\label{tab:performance_comparison}
\begin{tabular}{c|cc|cc|cc}
\toprule
\multirow{3}{*}{Method} & \multicolumn{2}{c|}{\multirow{2}{*}{MCL-JCI}} & \multicolumn{4}{c}{KonJND-1k} \\ \cline{4-7}
                        & \multicolumn{2}{c|}{} & \multicolumn{2}{c|}{JPEG} & \multicolumn{2}{c}{BPG} \\ 
\cline{2-7}
                        & $\Delta$JND & $\Delta$PSNR (dB) & $\Delta$JND & $\Delta$PSNR (dB) & $\Delta$JND & $\Delta$PSNR (dB) \\
\hline
PW-JND                  & 8.7      & 0.82       & 5.97       & 0.44        & 0.55       & 0.27 \\
SUR-Net                 & 5.22     & 0.63       & -          & -           & -          & - \\
SUR-FeatNet             & 4.44     & 0.58       & 6.95       & 0.50        & 1.46       & 0.76              \\
SG-JND                  & \textbf{4.26}     & \textbf{0.55}       & \textbf{5.39}       & \textbf{0.38}        & \textbf{0.45}       & \textbf{0.22}             \\
\bottomrule
\end{tabular}
\end{center}
\end{table*}

We first operate on the final feature vector \( F_{final}^{i} \) obtained from the feature extraction module to assess the perceptual quality of the whole distorted image. \( F_{final}^{i} \) is fed into two different branches, one to estimate patch quality score and the other to estimate the patch weight. Multiple fully connected layers regresses \( F_{final}^{i} \) to a patch quality score \( S_{patch}^{i} \):

\begin{eqnarray}
S_{patch}^{i} = MLP_1(F_{final}^{i}).
\end{eqnarray}

Another branch passes the feature \( F_{final}^{i} \) through another multiple fully connected layers, estimating the weight $W_{patch}^{i} $ of each patch:

\begin{eqnarray}
W_{patch}^{i} = MLP_2(F_{final}^{i}).
\end{eqnarray}

The overall quality score for the distorted image is then computed by combining the patch quality scores and weights:

\begin{eqnarray}
Q_{dist} = \sigma\left(\frac{\sum_{i=1}^{N} S_{patch}^{i} \cdot W_{patch}^{i}}{\sum_{i=1}^{N} W_{patch}^{i}}\right), 
\end{eqnarray}

\noindent
where \( \sigma \) is the sigmoid function, scaling the quality score to the range $[0, 1]$. Whether an image is perceptually lossy is determined by the corresponding binary label $L_{pred}$:

\begin{eqnarray}
L_{pred} = 
\begin{cases} 
1 & \text{if } Q_{dist} > 0.5,  \\
0 & \text{otherwise}.
\end{cases}
\end{eqnarray}

We consider this distorted image to be perceptually lossy when the label $L_{pred}$ is 1, and conversely we consider it to be perceptually indistinguishable from the reference image. The loss function used is the Binary Cross-Entropy (BCE) Loss between predicted labels $L_{pred}$ and ground truth labels $L_{gt}$.

With JPEG compression as a case study, for each given original image, we acquire a sequence of perceptual labels for each distorted image in the range of QF from 1 to 100, following the process delineated by the preceding modules.

The JND is defined by the boundary where the sequence of perceptual labels transitions from 0 to 1:

\begin{eqnarray}
JND = \min\{QF | L_{pred}^{qf} = 1\}. 
\end{eqnarray}

Acknowledging the possibility of random errors in neural network predictions, which may lead to incorrect jumps in label results, we employ a refined sliding window strategy \cite{liu2019PW-JND}. This method entails examining a contiguous set of QF and pinpointing the most consistent boundary where undetectable distortion becomes detectable. By methodically sliding this window across the QF spectrum and observing the labels within, we ascertain the more reliable prediction for the JND value.

The size of the sliding window is denoted as $w$, and the criteria for determining the JND is encapsulated by a threshold parameter $ \theta$. This is mathematically represented as:

\begin{eqnarray}
JND = \min\{QF_w | \sum _{qf=QF_w}^{QF_w+w} L_{pred}^{qf} \leq \theta\}. 
\end{eqnarray}

In this equation, $QF_w$ represents the starting QF of the sliding window. This robust strategy ensures that the JND prediction is not only based on the neural network’s output but also considers the inherent variability and potential anomalies in distortion perception, leading to a more robust and accurate JND prediction.

\begin{table*}[!h]
\centering
\captionsetup{font={footnotesize}}
\fontsize{8pt}{10pt}\selectfont
\caption{$\triangle JND$ OF FOUR MODELS ON MCL-JCI DATASET AND KonJND-1k DATASET FOR ABLATION STUDY.}
\label{tab:ablation_study}
\begin{tabular}{cccc|c|cc}
\toprule
\multirow{2}{*}{ResNet50} & \multirow{2}{*}{CSA} & \multirow{2}{*}{Patch Weight} & \multirow{2}{*}{search strategy} & \multirow{2}{*}{MCL-JCI} & \multicolumn{2}{c}{KonJND-1k}  \\ \cline{6-7}
                        & \multicolumn{3}{c}{}  & \multicolumn{1}{|c|}{} & JPEG & BPG \\ 
\hline
\checkmark &            &             &         & 11.82 & 9.1329 & 1.3313 \\
\checkmark & \checkmark &             &         & 5.28 & 5.7579 & 0.4563 \\
\checkmark & \checkmark & \checkmark  &         & 4.94 & 5.6131 & 0.4534 \\
\checkmark & \checkmark & \checkmark  & \checkmark  & 4.26 & 5.3869 & 0.4454 \\
\bottomrule
\end{tabular}
\end{table*}

\section{EXPERIMENT}

\subsection{Experiment Protocol}
\subsubsection{Dataset}
The proposed method is mainly validated on the MCL-JCI \cite{jin2016statistical} dataset and KonJND-1k \cite{lin2022large} dataset. The MCL-JCI dataset comprises 50 source images the resolution of 1920×1080, each associated with 100 JPEG-coded images with varying quality factors from 1 to 100. The KonJND-1k dataset includes 1,008 source images of 640×480 resolution, alongside distorted versions obtained using JPEG and BPG compression schemes which is the largest JND image dataset available.

\subsubsection{Evaluation Criteria}
We use two criteria to evaluate the performance of the proposed model and compared method, which are delta just noticeable distortion ($\triangle$JND) and delta peak signal-to-noise ratio ($\triangle$PSNR). $\triangle$JND measures the mean average error (MAE) between predicted and actual JND points, reflecting the model's accuracy in determining the perceptual threshold. $\triangle$PSNR assesses the MAE in PSNR values at the JND point, indicating the model's precision in capturing image quality changes at the JND point.

\subsubsection{Compared Methods}
We compared our proposed method with the following state of art models: PW-JND \cite{liu2019PW-JND}, SUR-Net\cite{fan2019net}, SUR-FeatNet \cite{lin2020featnet}.

\subsubsection{Experiment Setup}
For patch extraction, each input image is segmented into 16 patches of 64x64 pixels. For JND prediction, a sliding window of size 6 with a transition threshold of 5 is employed in JPEG compression format images, while a window of size 3 with a threshold of 2 is used in BMP compression format images. The proposed method is trained using parallel processing on two NVIDIA GTX 3090 GPUs. Optimization is conducted via the Adam optimizer with an initial learning rate of 0.0001 and a batch size of 16. The learning rate is reduced by 0.8 every 10 epochs, over 50 epochs for the the MCL-JCI and KonJND-1k dataset. JND samples from each source image were modeled using the generalized extreme value distribution \cite{lin2020featnet} to determine target JND values. We utilized 10-fold cross-validation to divide each dataset into 10 subsets, with 8 subsets used for training, 1 subset for validation, and 1 subset for testing in each experiment, and the epoch with the best performance on the validation subset was selected for testing each time. The results of these 10 tests are averaged to evaluate the performance.

\subsection{Performance Comparison}

The performance results on two JND datasets are shown in Table \ref{tab:performance_comparison}. The results reveal the superior proficiency of SG-JND in estimating JND values across the MCL-JCI and KonJND-1k datasets, and also indicate the robustness of SG-JND in handling different types of image distortions. Unlike PW-JND's direct feature extraction via convolution layers, SG-JND utilizes ResNet-50 as a backbone for pre-training to provide a more nuanced understanding of perceptual quality across various levels of detail, akin to SUR-Net and SUR-FeatNet. However, due to its efficient local distortion-focused method, SG-JND outperforms SUR-Net and SUR-FeatNet, especially in handling rich datasets. In addition, the modular architecture of SG-JND offers greater versatility in various image quality assessments, surpassing SUR-FeatNet's narrower focus on SUR curve prediction. The integration of semantic-guided feature fusion in SG-JND proves effective, surpassing other models in both accuracy and perceptual relevance.

Fig. \ref{histogram} and Fig. \ref{plcc} present a comprehensive analysis of the SG-JND's performance on the two datasets. For the MCL-JCI dataset (Fig. \ref{histogram}(a)), the absolute error in JND remains no more than 20 for all images, with the majority exhibiting an error of less than 10, accounting for $94 \%$ of the dataset. For the KonJND-1k dataset (Fig. \ref{histogram}(b)), the absolute error in JND is less than 10 for $83.5 \%$ of the JPEG compressed images and falls below 5 for $99.6 \%$ of the BMP compressed images. Fig. \ref{plcc} shows the Pearson linear correlation coefficient (PLCC) of the PSNR between the ground truth and predicted JND values across the three dataset types. The corresponding plcc values are 0.9746, 0.9866, 0.9719, all indicating high correlation. 

\subsection{Ablation Study}
In this section, we perform several ablation studies to further dissect the contributions of key components in SG-JND to validate their effectiveness in JND prediction. First, we build a baseline model that employs the lowest-level feature from the ResNet-50 backbone, obtains perceptual labels of distorted images through global average pooling and simple linear regression, and determines the JND value in the first transition from 0 to 1 in the set of perceptual labels.

The results are presented in Table \ref{tab:ablation_study}, commence with the baseline model's performance. Upon incrementally integrating the CSA mechanism, the patch weight module, and the sliding window search strategy, we observe successive enhancements in performance. Specifically, the inclusion of CSA significantly improves the semantic guidance of high-level features, while the patch weight module refines the model's focus on salient image regions. The sliding window search module further refines the JND prediction, mitigating abrupt transitions caused by potential neural network misclassifications.

Each component's addition demonstrates measurable benefits, with the full SG-JND configuration outperforming the baseline and individual ablated versions, confirming the hypothesis that both CSA and patch weighting, coupled with the sliding window strategy, are instrumental for superior JND prediction.

\section{CONCLUSION}
In this paper, we introduce SG-JND, an innovative method leveraging a deep neural network for precise JND prediction. SG-JND employs quality-focused pre-training for its neural backbone, enhancing its ability to detect subtle image distortions. In addition, it integrates an attention mechanism that utilizes high-level semantic features to guide the selection of low-level distortion-related features, which enriches the semantic analysis of the features. SG-JND demonstrates superior performance on two widely recognized JND datasets, outperforming current state-of-the-art methods. Our ablation study highlights the key role of attention in spatial feature analysis and the effectiveness of patch-weighting and sliding window methods in JND prediction. Future work could explore further optimization of these elements and their application in other domains of image processing.

\begin{spacing}{0.8}
\bibliographystyle{IEEEbib}
\bibliography{refs}
\end{spacing}

\end{document}